\documentclass[conference]{IEEEtran}
\IEEEoverridecommandlockouts
\usepackage{cite}
\usepackage{amsmath,amssymb,amsfonts}
\usepackage{algorithmic}
\usepackage{graphicx}
\usepackage{textcomp}
\usepackage{xcolor}
\usepackage[letterpaper,top=0.75in,right=0.63in,left=0.63in,bottom=1.04in]{geometry}

\def\BibTeX{{\rm B\kern-.05em{\sc i\kern-.025em b}\kern-.08em
    T\kern-.1667em\lower.7ex\hbox{E}\kern-.125emX}}
    
\begin{document}

\title{Digital Sovereignty Control Framework for Military AI-based Cyber Security}

\author{
     \IEEEauthorblockN{Clara Maathuis\IEEEauthorrefmark{1}, Kasper Cools\IEEEauthorrefmark{2}\IEEEauthorrefmark{3}
    }
    
    \IEEEauthorblockA{
        \IEEEauthorrefmark{1}Open University of the Netherlands\\
        \IEEEauthorrefmark{2}Royal Military Academy, Belgium\\
        \IEEEauthorrefmark{3}Vrije Universiteit Brussel, Belgium\\
        \IEEEauthorrefmark{1}clara.maathuis@ou.nl, 
                 \IEEEauthorrefmark{2}kasper.cools@mil.be 
    } 
}

\maketitle

\begin{abstract}  
In today’s evolving threat landscape, ensuring digital sovereignty has become mandatory for military organizations, especially given their increased development and investment in AI-driven cyber security solutions. To this end, a multi-angled framework is proposed in this article in order to define and assess digital sovereign control of data and AI-based models for military cyber security. This framework focuses on aspects such as context, autonomy, stakeholder involvement, and mitigation of risks in this domain. Grounded on the concepts of digital sovereignty and data sovereignty, the framework aims to protect sensitive defence assets against threats such as unauthorized access, ransomware, and supply-chain attacks. This approach reflects the multifaceted nature of digital sovereignty by preserving operational autonomy, assuring security and safety, securing privacy, and fostering ethical compliance of both military systems and decision-makers. At the same time, the framework addresses interoperability challenges among allied forces, strategic and legal considerations, and the integration of emerging technologies by considering a multidisciplinary approach that enhances the resilience and preservation of control over (critical) digital assets. This is done by adopting a design oriented research where systematic literature review is merged with critical thinking and analysis of field incidents in order to assure the effectivity and realism of the framework proposed.
\end{abstract}


\section{Introduction\label{section:introduction}}
Traditionally linked to the protection of physical territories and political authority, sovereignty in the digital era extends to safeguarding information systems, critical infrastructure, and communications networks against cyber threats that could alter a state's ability to exercise independent control~\cite{CyberSovereignty}. In this context, the meaning of sovereignty has diversified: it encompasses the defense of national digital assets from external attacks, the jurisdictional assertion over cyber incidents, and the broader imperative of ensuring that technological deployments do not undermine state authority or democratic principles becoming essential maintaining political, social, and technical stability \cite{DigitalSovereigntyPolitics}. As states see to (re)assert control not only over physical territory but also over digital infrastructures, networks, and information systems, they aim to maintain their authority and resilience in this highly interconnected world \cite{DigitalSovereigntyStakeholders}. This evolving control mechanism that finds itself under the umbrella of digital sovereignty, reflects the need to manage and safeguard the physical resources, software architectures, and information flows that shape digital realities \cite{DigitalSovereigntyDefinition}. While digital sovereignty initially emerged as a defensive response to external technological dependencies and vulnerabilities, it has since become a strategic objective across a wide range of political regimes for various stakeholders, from authoritarian models of territorial closure to liberal efforts emphasizing infrastructural resilience and regulatory leadership \cite{DigitalSovereigntyBuzz}. Nevertheless, digital sovereignty remains an ambiguous concept -- even as its use becomes more widespread -- with varying interpretations depending on political, cultural, and strategic contexts. States seldom define sovereignty explicitly in cyber security or digital policy strategies, and where it is invoked, the term often lacks consistency and operational clarity \cite{CyberSovereignty}. The core idea generally revolves around control, but this foundational premise is layered with divergent normative and geopolitical interpretations, ranging from national security imperatives to geoeconomic strategies for global standard setting \cite{DigitalSovereigntyDefinition, DigitalSovereigntyBuzz}. Moreover, the use of the term as a political buzzword often obscures underlying tensions between national autonomy and global digital interdependence, complicating coherent system design and governance mechanisms that need to be put in place \cite{DigitalSovereigntyBuzz}.

As a concept, digital sovereignty finds itself at the intersection of technology, governance, and societal values, reflecting the increasing need for individuals and states to exert control over their digital infrastructures, data, and technological development. Defined by the EU Federal Chancellery as the ability to shape digital transformation in a self-determined manner regarding hardware, software, services, and skills, digital sovereignty emphasizes making sovereign decisions within a legal framework without resorting to isolationism. Beyond its technical meaning, digital sovereignty has political and economic ramifications, as states seek to reassert authority over digital infrastructures increasingly dominated by private tech giants such as Google, Apple, Meta, Amazon, and Microsoft \cite{DigitalSovereigntyDefinitionC}. This ambition is particularly visible in the European Union’s (EU) efforts to channel regulatory initiatives aimed at curbing platform dominance and defending European strategic autonomy, highlighting the EU’s vision of becoming a "regulatory superpower" in the global digital order \cite{DigitalSovereigntyDefinitionC, DigitalSovereigntyEU}. However, this ambition faces tensions between aspirations for territorial control and the realities of global interconnectivity, as digital networks transcend national borders \cite{DigitalSovereigntyEU}. Moreover, digital sovereignty discourse often oscillates between defensive protectionism and proactive standard-setting, revealing a dual strategy and perspective: both securing internal digital assets and projecting influence globally through regulatory leadership \cite{DigitalSovereigntyBuzz}. Hence, building and achieving digital sovereignty requires not only technical expertise and regulatory assertiveness, but also a critical understanding of its layered meanings and practical implications across political, economic, and societal domains that needs to be assured at both national and individual levels \cite{DataSovereigntyBlurry}. In these lines, cyber sovereignty concerns the application of established principles of state sovereignty such as territorial integrity and non-intervention to cyberspace \cite{CyberSovereignty}. In cyberspace, states strive to assert authority over digital activities crossing their borders. A key element underpinning both digital and cyber sovereignty is the control over data, which represents the strategic resource of the information age. The criticality of data stems from its ability to confer power and influence, affecting everything from national security to individual privacy. Nevertheless, classic cyber security measures often fail to prevent unauthorized access to data or dissemination of data once it is accessed, highlighting significant security issues that need to be dealt with \cite{DataSovereigntyCon}.

Hence, data sovereignty is a rich and multidimensional notion that encompasses technical, legal, societal, ethical, and cultural dimensions. It directly refers to meaningful control, ownership, and stewardship of data, often expressed by various stakeholders across contexts including legislation, ICT architectures, and societal advocacy \cite{DataSovereignty}. However, as recent studies show, the term is used with considerable variance and should not be used as a synonym to digital sovereignty, given their different focal points and implications \cite{DataSovereigntyBlurry}. Beyond technical aspects, data sovereignty intersects with cultural and ethical values, emphasizing the importance of inclusivity, respect for societal norms, and promotion of local innovation and economic development \cite{DataSovereigntyCulture}. Thus, the challenge lies in operationalising data sovereignty in a way that acknowledges its layered meaning.  

Data sovereignty ensures that individuals, communities, organizations, and states maintain authority over their digital assets, yet when data is used to build AI models, new sovereignty challenges arise. The shift from human-based to AI-based or assisted decision-making processes risks systematically devaluing human judgment and transferring discretionary power to non-human agents. In parallel, the concentration of AI development and deployment capabilities within a few dominant technology firms leads to concerns that these entities now exercise a form of digital sovereignty as they shape the control from traditional state actors. Thus, achieving digital sovereignty when building and using AI-based systems requires the stakeholders involved to not only secure the data used by these systems, but also to govern and control the lifecycle, autonomy, and strategic deployment of the AI models themselves \cite{AIGovernance}.

This implies that states and organizations lacking effective digital sovereignty frameworks risk losing control 	foreign or corporate actors. Such a loss can undermine their technological autonomy, economic competitiveness, and societal resilience. Thus, the development of tailored sovereignty mechanisms that balance freedom, regulation, and technological development is crucial to securing strategic advantages across all domains, and in particular in the military cyber domain. Military forces increasingly recognize that cyberspace is a battlefield requiring sovereign capabilities for both defensive and offensive digital operations \cite{DigitalSovereigntyMilitary}. In this context, systems need to be designed to secure curated defence data, ensure assertive management of suppliers and contracts, and maintain sovereign control over AI models and infrastructures \cite{DigitalSovereigntyUKMoD}. At the same time, AI-based systems developed and deployed for cyber defense and offense purposes amplify both opportunities and vulnerabilities: their ability to act with high autonomy or autonomously necessitates mechanisms ensuring that strategic decisions remain aligned with national interests and legal frameworks. Moreover, given the fluidity and complexity of cyber threats that range information manipulation to critical infrastructure attacks \cite{DigitalSovereigntyModels}, digital sovereignty in military cyber operations must emphasize resilient, adaptable, and verifiable control frameworks. In doing so, armed forces not only protect their cyber capabilities, but also contribute to shaping a future international order where sovereignty in cyberspace is respected and enforced through credible, responsible, and trustworthy action. 

In this context, this research aims to design a framework for digital sovereign control of military AI-based cyber security systems grounded in a systematic and multidisciplinary approach. Specifically, the framework is built upon two core methodological pillars. First, a systematic literature review for identifying and analysing existing datasets used in military AI-based cyber security applications, thus providing a concrete understanding of the empirical resources used by the AI systems in this domain. And second,  an extensive literature review examining the meaning, evolution, and implications of digital sovereignty across various domains and in particular in the military domain, is carried out. To the best of our knowledge, this represents the first effort in this direction. Through this dual research strategy, it is ensured that the proposed framework is both practically anchored in real-world data practices and conceptually robust in addressing the multidimensional challenges of sovereignty in this domain. For demonstration purposes, a use case is presented for the framework proposed on cyber operations and efforts on this behalf carried out in the ongoing war in Ukraine. This case study provides critical insights into the challenges, dynamics, and imperatives of maintaining sovereignty over digital infrastructures, data, and AI models under conditions of active geopolitical confrontation.

The outline of this paper is structured as follows. Section~\ref{section:related-work} discusses related studies carried out in this context. Section~\ref{section:methodology} presents the methodological approach considered in this research. Section~\ref{section:datasets} reflects on the categories of datasets used when building AI-based cyber security applications in the military domain. Section~\ref{section:framework-design} proposes the design of the framework advanced in this research. Section~\ref{section:framework-use-case} demonstrates the applicability of the framework proposed on a use case conducted on various cyber operations carried during the ongoing war in Ukraine. At the end, Section~\ref{section:discussion-conclusions} discusses concluding remarks and future research perspectives. 

\section{Related work\label{section:related-work}}
The integration of digital technologies in military operations has led to the development of multi-domain operations, which prioritise the alignment of activities across all operational domains, including cyberspace~\cite{MultiDomain}. This method is designed to achieve unified results quickly, confronting the multifaceted and nonlinear aspects of current warfare. The ongoing war in Ukraine has highlighted the important role of cyber operations in modern warfare, demonstrating the need for resilient and adaptable control frameworks~\cite{ContestingSovereigntyCyberSpace,CyberWarfareComm}.

In the face of escalating tensions, digital sovereignty is essential. States strive to master information technologies and maintain sovereign digital spaces while being part of a global network. This involves regulating cross-border data flows for security reasons and coordinating cybersecurity regulations globally~\cite{ControllingContentEU, DigitalSovereigntyProtectionismAutonomy}. The challenges include state-sponsored cyber operations, cyber espionage, and the dynamic environment shaped by advanced technologies like AI and quantum computing. The Russia-Ukraine conflict has underscored the importance of cyber resilience and the strategic use of cyber operations to protect national interests~\cite{BeyondSovereignty}. \cite{CyberDigitalSovereigntyTNO} advocate for a framework for cybersecurity sovereignty action encompassing relevant policy domains and instruments. The authors emphasize the importance of being able to assess possible improvements of capabilities, capacities, and control. They also highlight the importance of increasing these elements in cybersecurity sovereignty. \cite{DigitalSovereigntyChallengesAndPerspectives} highlights that digital sovereignty is a relatively new concept, still evolving without a formal definition. The research emphasizes the necessity of considering digital sovereignty while ensuring that organizations and states maintain a balance, avoiding restrictions on their foreign economic relations. Frameworks must accommodate multiple sovereign interests and adapt to this balance.

Data sovereignty refers to the control over data storage, transfer, and use. Countries are increasingly implementing data localization laws and regulations to safeguard national interests. The European Union, for instance, has established the Digital Markets Act and the Digital Services Act to regulate the digital economy and emerging technologies. These measures aim to foster homegrown tech industries and ensure technological independence, deepening geopolitical competition. The control over data is a strategic resource in the information age, influencing everything from national security to individual privacy~\cite{BeyondControlData}. \cite{DataSovereignntyGovernanceFramework} propose a data sovereignty governance framework using knowledge graphs for data classification. This work emphasizes the importance of being able to identify and classify data in order to comply with regulations. \cite{DataSovereigntyIS} introduces a conceptual model for understanding data sovereignty which encompasses seven core aspects needed to ensure data sovereignty and guarantee trusted data sharing between different parties.

Sovereign AI involves the control over AI technologies to preserve national values, privacy, and security. This concept extends the core components of digital sovereignty to AI, adding a value alignment component. Governments are developing strategies to achieve sovereign AI, focusing on domestic capabilities and control. The interplay between AI autonomy and control is critical, with varied consequence depending on the level of control applied. The integration of AI in military operations, particularly in cyber defense and offense, amplifies both opportunities and vulnerabilities~\cite{DigitalSovereigntyTimeOfConflict, DigitalSovereigntyProtectionismAutonomy}.

Although a lot has already been done on a regulatory basis, existing literature has yet to comprehensively address the intersection of digital sovereignty with AI-based cyber security systems in the military domain. Addressing the multidimensional challenges of sovereignty in this field is essential. This research highlights the importance of a tangible framework within the domain of digital sovereignty for effectively managing modern military operations, cybersecurity, data management, and AI governance within the defense sector.

\section{Research methodology\label{section:methodology}}
This research aims to design a Digital Sovereign Control framework for military AI-based cyber security systems. On this behalf, the following research questions are formulated: 
\begin{itemize}[noitemsep]
\item Which datasets or data sources are used to develop AI-based cyber security applications in the military domain?
\item What are the key elements for designing a Digital Sovereignty Control framework for these applications in this domain?
\end{itemize} 

To tackle these research questions, a multidisciplinary stance is adopted in two phases. In the first phase, a Systematic Literature Review (SLR)~\cite{PRISMA} is conducted for meticulously identify, analyse, and categorize the diverse datasets essential for developing, training, and deploying AI systems in military cyber contexts, encompassing areas such as cyber defence, cyber warfare, blockchain analysis, social media intelligence, agent-based modelling, and specialized training corpora. This structured approach provided critical foundational knowledge regarding the specific data artefacts over which sovereignty must be asserted and maintained, reflecting the inherently multidisciplinary nature of the problem which intersects AI, cyber security, military operations, and policy studies. In this phase, the search string includes keywords like cyber security, cyber warfare, military, defence, artificial intelligence, machine learning, and deep learning. Accordingly, the string was used to conduct searches on scientific databases such as EEE Digital Library, ACM Digital Library, Web of Science, Taylor \& Francis, Sage Journals, Wiley, and Google Scholar (first 10 pages). This resulted in initial 3860 articles from which 52 articles resulted at the end for being analysed in-depth. 

In the second phase, building upon the empirical grounding provided by the SLR, an extensive literature review focused specifically on the concept of digital sovereignty itself inside the same scientific databases. This review examined existing theoretical frameworks, conceptual models, relevant doctrines, evolving policy considerations, and technological enablers pertinent to establishing and verifying sovereign control over digital resources and processes, both generally and within security-sensitive contexts from an initial set of 51 articles. In addition to these resources, an analysis on various cyber operations executed during the war in Ukraine was applied. By merging the practical data landscape mapped by the SLR with the theoretical and use case insights gathered from the extensive review of digital sovereignty and cyber operations conducted in Ukraine, these resources combined in a multidisciplinary research approach allowed the informed design of the proposed framework. This ensures the resulting framework is not only conceptually robust, addressing core sovereignty principles, but also practically relevant to the specific data and operational realities encountered in military AI-driven cyber security.

\section{Domain datasets\label{section:datasets}}
Understanding which categories of datasets are utilized for developing AI-based cyber security systems in the military domain is essential for effective digital sovereignty control, as it informs strategic alignment, threat mitigation, and operational effectiveness. This necessity arises from the inherent complexity, diversity, and uncertainty that surrounds cyberspace entities and incidents occurring inside or through cyberspace. To this end, based on the systematic literature review conducted, seven categories of datasets were identified: cyber defence, cyber warfare, blockchain, social media, Unmanned Aerial Vehicle (UAV), agent-based modelling and simulation, and learning and ethics. While cyber defence and warfare datasets support real-time threat detection and proactive defensive actions, blockchain datasets ensure the assessment of secure and transparent data sharing, social media datasets enable the identification and mitigation of disinformation threats, agent-based modelling datasets facilitate predictive scenario analysis, and education datasets enhance cybersecurity awareness and preparedness. The following overview explicitly identifies and outlines these dataset categories, emphasizing their strategic importance and distinct roles in the broader context of military digital sovereignty and cyber operations.

In the first category, Cyber Defence, the Simulated Control System Communications Dataset \cite{P2} is used to simulate military command and control interactions allowing algorithms to spot protocol anomalies in real time. Complementary malware corpora such as the VirusTotal Dataset \cite{P9, P16} ] support IoT oriented detection pipelines, while the Markamilley.com Dataset \cite{P11} illustrates how misinformation attacks are conducted on senior leadership in order to contaminate the cyber terrain. At the same time, the NATO Data Farming Services Dataset \cite{P23} structures experiment run metadata for risk–benefit assessment of AI concepts, and the MIL STD 1553 Communication Bus Dataset \cite{P30} allows protocol level defence research, while the comprehensive Zeek-based Cyber Datasets \cite{P28}, combining generated and synthetic data, enable AI models to learn complex network traffic and user behaviour patterns for early threat detection.

In the second category, Cyber Warfare, the Cyber-Warfare Event Compatibility Dataset \cite{P4} uses a broad spectrum of scenarios, famously including the Stuxnet attack, to train AI models. The goal is to equip these models to recognize and effectively counteract sophisticated adversary tactics, techniques, and procedures. Alongside this, the Cyber Attack Simulation Data \cite{P18} provides a complementary resource. It consists of artificially generated cyber attack scenarios designed to mirror real-world threats. This incorporates computational data like memory and CPU usage, enabling military forces to evaluate how well their systems withstand and perform under simulated duress.

In the third category, Blockchain, the Blockchain Military Dataset \cite{P5}, containing transaction, smart contract, and interaction data, is valuable for developing AI models designed to secure military blockchain applications against unauthorised access and hacking, facilitating automated security protocols and real-time anomaly detection in transaction data.

In the fourth category, Social Media, the Markamilley.com Dataset \cite{P6} provides concrete examples of how social media platforms can be exploited to damage reputations and propagate misinformation through techniques like fake messages and manipulated images. To understand the reception of military actions, the Sentiment Analysis Dataset [29] offers capabilities for gauging public reactions and sentiments. Within specific geopolitical contexts, such as the ongoing war in Ukraine, the Twitter Sentiment Dataset [30] provides a deeper understanding of competing narratives by classifying sentiments, e.g., differentiating between pro- and anti-Kremlin viewpoints. Offering a more integrated perspective, the Ukraine Conflict Datasets \cite{P40} correlate official communications, such as speech transcripts from President Zelensky, with reported violent incidents. This structured approach yields valuable insights into wartime public communication strategies and the associated emotional responses they elicit.

In the fifth category, Unmanned Aerial Vehicles (UAV), the UAV Security Dataset \cite{P8, P12} developed in a synthetic testbed, is tailored for evaluating machine learning-based anomaly detection for drones. Image analysis capabilities are supported by the SatUAV dataset \cite{P21} containing satellite-aerial image pairs for identifying discrepancies. At the same time, standard benchmarks are integrated with drone-specific features such as GPS data in datasets like the LRRF Model Datasets \cite{P27} and the Merged Drone Dataset \cite{P29}. Moreover, scenario-specific resources like the Attack Tree Model Dataset \cite{P31} model threats such as surveillance data theft, while the UAV-Assisted HetNet Security Dataset \cite{P45} focuses on identity-based authentication in complex network environments involving UAVs and small cells.

In the sixth category, Agent-based modelling and Simulation, the datasets facilitate research into autonomous systems and complex interactions. On this behalf, the Autonomous Intelligent Cyber-defense Agent (AICA) Dataset \cite{P41} supports exploration into agent functions like sensing, decision-making, and action selection within cyber security. The Red-Blue Team Cyber Interaction Dataset \cite{P42} captures the complexities of attack-defence scenarios, providing high-level descriptions and detailed network models. Furthermore, the Adaptive C2 Network Topology Dataset \cite{P43} uses generated network structures to simulate threats against Command and Control networks, enhancing understanding of communication resilience.

In the seventh category, Learning and Ethics, the Cyber Gym for Intelligent Learning (CyGIL) dataset \cite{P7} provides a distinctive environment, combining an emulator and a high-fidelity simulator. This setup allows AI agents to be trained using realistic reinforcement learning scenarios specifically designed for cyber operations. Understanding the baseline for normal system operations, which is essential for effective learning and anomaly detection, is supported by the System Operational Data dataset [19]. This captures the signals transmitted within military systems and the data stored in real-time memory. Furthermore, the ethical dimension of military AI is addressed by datasets such as the Ethics of GLAWS \cite{P22}. This particular dataset captures the outcomes from a United States Military Academy pilot program where students applied Just War Theory principles to robot programming, thereby fostering a necessary and deeper understanding of the moral implications inherent in deploying AI within military contexts.

\section{Framework Design\label{section:framework-design}}

Following the delineation and categorization of essential datasets used for military AI-based cyber security applications, considerations are given to understanding what digital sovereignty means and what digital sovereignty control implies in this context. This requires extending verifiable control beyond the raw data artefacts – encompassing secure lifecycle management, jurisdictional oversight, and stringent access protocols – to the AI models built from them. Assuring dual sovereignty over both data and AI models is fundamental to maintaining operational integrity, safeguarding sensitive defence information, preventing adversarial manipulation, and ensuring that AI-based cyber security systems reliably support national strategic imperatives and ethical considerations in the military domain. 

Based on the systematic literature review that identified the diverse datasets employed in AI-based military cyber security applications and simultaneously the extensive literature review conducted on digital sovereignty within this domain and beyond, the design of a conceptual framework for Digital Sovereignty Control is further proposed. Specifically tailored for military AI-based cyber security contexts, this framework is structured around four essential, interconnected elements deemed critical for effective governance, as follows. Firstly, context, which defines the strategic, legal, and ethical operating boundaries. Secondly, autonomy, which addresses the control over AI decision-making and operational independence. Thirdly, stakeholder Involvement, managing the human and organizational interfaces, including allied collaboration and oversight. And fourthly, mitigation of risks, ensuring the security, resilience, and trustworthiness of systems against relevant threats. These four pillars collectively aim to provide a comprehensive structure for establishing, assessing, and maintaining verifiable sovereign control over critical digital assets and AI-driven processes in this critical domain. 

\subsection{Framework Layers\label{section:framework-layers}}
The framework proposed has a layered architecture that ranges from high-level strategic direction down to operational execution and technical assurance. 

The Strategic Layer represents the base of the framework as it is establishing the overarching rules and mandates in this domain. This implies deciding policy, legal, and ethical considerations that need to be applied to assure digital sovereignty in this domain. At the same time, the transparency of the solutions built and used need to be ensured through concrete directives that assure interpretable outcomes for accountability purposes and promotion of integration and respect of corresponding legal, social, and ethical principles, norms, and values aligned with relevant doctrines and rules that  are shaped by actors such as the Ministry of Defence policy office, legal counsel, data protection authorities, and alliance representatives, and enforced through relevant mechanisms \cite{AIGovernance, CyberSovereignty, DataSovereignty, DigitalSovereigntyUKMoD, DigitalSovereigntyEU, DigitalSovereigntyDefinitionC} . 

The Governance and Assurance Layer focuses on oversight, detailed policy formulation, and continuous verification of digital sovereignty. This layer utilizes tools such as sovereign maturity metrics and red-teaming exercises, audits, immutable trails, and ethics review boards to ensure compliance and trustworthiness. Furthermore, in this layer translates the high-level strategy goals into specific actionable policies by defining the National Digital Sovereignty Doctrine in the military cyber context, including control over data, algorithms, and infrastructure for defence and offence while assuring level compliance with frameworks such as IHL (International Humanitarian Law), data protection mechanisms such as DPIA (Data Protection Impact Assessment), and RoE (Rules of Engagement) while implementing robust data governance policies that assure the definition of corresponding sensitive levels and lifecycle handling procedures \cite{DataSovereigntyBlurry, DigitalSovereigntyBuzz, DataSovereigntyCulture, DigitalSovereigntyModels, DigitalSovereigntyMilitary, USDigitalSovereignty}. 

The Data Sovereignty Layer reflects the need for verifiable control over information assets throughout their lifecycle. This involves determining jurisdictional and operational control, establishing clear classification and handling rules, and employing technical measures such as crypto-sharding, confidential computing, and federated learning gateways. In this sense, protection controls are tailored per dataset category, examples including real-time streaming of cyber operations logs, employing data-diodes and on-premise vector databases for Social-Media OSINT to prevent exfiltration, using hashing for Agent-Based Simulations to ensure reproducibility, and implementing graded de-classification for education or training corpora. This layer ensures data residency in sovereign storage, secures data transit, and applies mechanisms such as data minimization \cite{DataSovereigntyBlurry, DataSovereigntyCon, DataCyberSovereignty, DataSovereignty, DataSovereigntyCulture, DigitalSovereigntyExample, DigitalSovereigntyAutonomy}. 

The AI Sovereignty Layer refers to the AI models used by requiring secure MLOps pipelines, model provenance ledgers comprehensive adversarial and robustness testing, and policy-aware model deployment. Such techniques are federated or split learning pipelines that keep raw training data within secure environments while allowing model updates across domains, and dedicated model verification and validation together with red teaming ranges for adversarial testing, fairness assessment and bias scanning, and mission-task fitness evaluation. In this layer, considerations are given to building transparent and interpretable AI using various relevant Explainable AI (XAI) methods and techniques \cite{AIGovernance, DataSovereignty, DigitalSovereigntyUKMoD, DigitalSovereigntyAutonomy}. 

The Cyber Operations and Infrastructure Layer ensures sovereign control during the practical application and execution of AI-driven cyber capabilities, encompassing both the operational procedures and the underlying technological infrastructure. This includes managing defensive AI (e.g., anomaly detection), offensive cyber planning support, autonomous agents, and simulations, while actively identifying and mitigating digital sovereignty threats like unauthorized access and supply-chain attacks. Further this implies using trusted hardware and software through prioritizing nationally developed or rigorously vetted components and secure processes, maintaining a sovereign computing environment using national military clouds, and ensuring network infrastructure control over military communication networks using principles such as security and privacy by design, zero-trust access, mechanisms for live human override (i.e., human on the loop), and battle damage assessment capabilities \cite{AIGovernance, DataSovereigntyCon, CyberSovereignty, DataCyberSovereignty, DataSovereigntyCulture, DigitalSovereigntyUKMoD, DigitalSovereigntyEU, DigitalSovereigntyMilitary, DigitalSovereigntyAutonomy, DigitalSovereigntyDefinitionC, USDigitalSovereignty}. 

\subsection{Framework Dimensions\label{section:framework-dimensions}}
The dimensions that characterize this framework are defined as follows: 

The Context dimension refers to a comprehensive understanding of the operational environment within which these military AI-based cyber security systems function. This involves identifying and classifying mission-critical datasets, rigorously mapping the threat landscape, including potential cyber threats like unauthorized access, ransomware, supply-chain compromises, disinformation campaigns, espionage, and hybrid warfare scenarios. At the same time, it also relates to the jurisdictional, regulatory, strategic, and geopolitical boundaries that assure that data storage, processing, and execution comply strictly with national and international legal and socio-ethical standards and principles. Furthermore, it assures the use of dedicated sovereign infrastructure to provide robust physical and digital control over sensitive military data and systems accounting both the requirements and goals of all stakeholders involved \cite{AIGovernance, DataCyberSovereignty, DataSovereignty, DataSovereigntyCulture, DigitalSovereigntyUKMoD, DigitalSovereigntyModels, DigitalSovereigntyEU}.

The Autonomy dimension addresses the controlled independence granted to AI systems and the operational resilience required to sustain them, ensuring alignment with national sovereignty objectives. It involves clearly delineating the decision-making capabilities of AI within the military context, establishing a careful balance between AI-based processes and requisite human oversight. This refers to the (i) operational autonomy which requires the establishment of independent, resilient infrastructures capable of maintaining uninterrupted AI services even when subjected to cyber attacks or heightened geopolitical tensions, and (ii) policy-driven autonomy governed by explicit, digitally embedded rules, operational constraints, and predefined conditions, ensuring that AI actions remain bounded and serve national interests while preserving the ultimate decision authority of human commanders/decision-makers. Through this dimension, autonomy tiers are defined and used (e.g., human-in-the-loop, human-on-the-loop, full autonomy) reflecting operational context and sensitivity, governed by specific autonomy governance policies and protocols that establish when AI can act independently versus requiring human approval or in human-AI teaming settings considering an adaptive autonomy approach for dynamically adjustment and self-healing controls based on situational risk assessments, context awareness, and system confidence levels. At the same time, (real-time) monitoring and intervention need to be considered to assure the transparency of the systems used by means of dashboards and integration of explainability mechanisms \cite{AIGovernance, DataCyberSovereignty, DataSovereigntyCulture, DigitalSovereigntyUKMoD, DigitalSovereigntyEU, DigitalSovereigntyMilitary, DigitalSovereigntyAutonomy, DigitalSovereigntyDefinitionC}.

The Stakeholder Involvement dimension captures the defining roles that the stakeholders involved have, their collaboration mechanisms while implementing appropriate oversight, and ensuring adequate training and awareness. This acknowledges the necessity of involving a wide array of actors, ranging from strategic leadership (e.g., Ministers of Defence, policy boards) and operational commanders (e.g., joint task force commanders, cyber security/defence centres) to technical personnel (e.g., AI engineers, data stewards, cyber operators), legal and ethics advisors, intelligence communities, allied partners, industry and academia involved in the supply chain or research, and ultimately, the end-users and trainers interacting with the systems. The core purpose is to embed sovereignty considerations within the established military command structure while facilitating necessary cooperation and ensuring accountability across all levels \cite{DataSovereigntyBlurry, DataSovereignty, DigitalSovereigntyModels, DigitalSovereigntyAutonomy}.

The Risk Mitigation dimension implies proactively identifying, analyzing, prioritizing, and neutralizing threats that could compromise digital sovereignty, operational continuity, system security, and trustworthiness within the military AI cyber domain. This encompasses a broad spectrum of risks, including specific AI vulnerabilities (e.g., model poisoning, evasion, extraction), traditional cyber threats (such as unauthorized access and ransomware), critical infrastructure weaknesses, insider threats, ethical and legal non-compliance issues, and particularly pernicious threats to sovereignty like supply-chain attacks. This dimension requires a systematic and continuous approach by employing comprehensive risk modelling and assessment tailored to each dataset category and system component ensuring that potential attack vectors and their corresponding exploited vulnerabilities are understood and addressed before they can even be exploited, thereby preserving control. On this behalf, robust technical and procedural safeguards need to be applied considering a layered defence mechanism, resilient recovery methods, and active threat monitoring and testing solutions \cite{CyberSovereignty, DataCyberSovereignty, DataSovereignty, DigitalSovereigntyExample, DigitalSovereigntyDefinitionC, DigitalSovereigntyDefinitionC}.

\subsection{Framework Metrics\label{section:framework-metrics}}
The successful implementation and assessment of the proposed framework implies assessing its effectiveness using applicable metrics in relevant scenarios. Without objective metrics, assessments of sovereignty risk becoming subjective, hindering systematic evaluation, identification of vulnerabilities, justification of resources, demonstration of compliance with legal and ethical mandates, and the tracking of improvements over time. In this sense, the metrics serve to operationalize the abstract concept of digital sovereignty, transforming principles like data localization, policy-bound autonomy, secure interoperability, and supply-chain integrity into quantifiable indicators, thereby providing demonstrable assurance critical for military AI cyber security systems. Accordingly, various control metrics should be aligned with the framework's structure and objectives when applied to specific use cases. Drawing from the detailed components discussed, potential metrics could include \cite{AIGovernance, DigitalSovereigntyMilitary, DigitalSovereigntyModels, DigitalSovereigntyAutonomy, DigitalSovereigntyDefinitionC, USDigitalSovereignty, DigitalSovereigntyModels, DigitalSovereigntyMilitary, DigitalSovereigntyNor}: 

\begin{itemize}[noitemsep]
\item Compliance Metrics such as percentage adherence to national/IHL/ethical guidelines, compliance with governance bodies like the EU AI Act and EU Cyber Resilience act, audit trail completeness from blockchain logs, and policy-as-code enforcement accuracy.
\item Autonomy Control Metrics such as frequency/latency of human overrides, validation rates for trust, and  adherence to defined autonomy tiers.
\item Data and Model Sovereignty Metrics such as percentage of critical data/models with verified provenance via a mechanism such as SBOM (Software Bill of Materials) validation, data exfiltration prevention success rate via data diodes, and attack detection rates.
\item Resilience Metrics such as infrastructure uptime in sovereign environments, success rates of fail-safe/degradation protocols, and vulnerability remediation times for critical components.
\item Sovereign Control Index to assess the degree to which digital sovereignty is being effectively maintained for a specific military AI-based cyber security system or capability, e.g., effectiveness of access controls and assurance level of model provenance. 
\item Sovereignty Maturity Metric to assess the level of capability, effectiveness, and sophistication with which an organization or specific system implements and manages digital sovereignty controls using a graduated scale with several defined levels. This serves as a standardized way to benchmark the current state of digital sovereignty, identify areas for improvement, define target maturity levels based on risk and strategic importance, and track progress over time in this domain. 
\end{itemize}

\section{Use Case\label{section:framework-use-case}}
The ongoing war in Ukraine is considered as a use case to demonstrate and assess the framework proposed given the complexity and intensity of the cyber operations conducted in this context. This use case specifically examines both the layers and dimensions of the framework based on lessons learned from real operational contexts, focusing on various incidents such as cyber attacks on Ukrainian critical infrastructure, disinformation campaigns, and sophisticated supply-chain attacks. Specifically, these cyber operations have often been integrated with conventional military actions, aiming to disrupt Ukrainian command and control, degrade critical infrastructure, conduct espionage, and undermine societal morale and government legitimacy \cite{mankoff2024ukraine} The strategic intent appears multifaceted, ranging from direct tactical support for kinetic operations (e.g., disrupting communications ahead of an advance) to broader strategic efforts aimed at destabilizing the Ukrainian state and influencing international perceptions. At the same time, social media manipulation campaigns such as the ones based on disinformation have been a pervasive element of the conflict. State-sponsored actors and proxies have utilized social media platforms, state-controlled media, and covert online operations to spread narratives aimed at justifying the invasion, discrediting Ukrainian leadership, building division within Ukraine and among its allies, and potentially inciting unrest These incidents highlight the urgent need for robust digital sovereignty measures capable of maintaining operational autonomy, security, and ethical alignment when countering various hybrid warfare tactics.

Despite the intensity and sophistication of these cyber attacks, Ukrainian cyber defenses have demonstrated remarkable resilience, significantly bolstered by international support from both governments and the private sector. Ukraine's Computer Emergency Response Team (CERT-UA) has been effective in identifying threats and coordinating responses, often publicly sharing technical details of attacks \cite{UseCase11}. At the same time, major technology companies like Microsoft and Google have provided extensive threat intelligence, cyber security support, and cloud infrastructure resilience, playing a crucial role in mitigating attacks and maintaining essential services. Furthermore, intelligence sharing and capacity-building efforts from allied nations have significantly enhanced Ukraine's ability to anticipate, detect, and respond to cyber threats in near real-time, representing an unprecedented level of international cyber defense cooperation during active conflict. In this context, clear roles are defied, secure collaboration frameworks respecting sovereignty caveats are formulated, rapid training on diverse tools are created and executed, and cross-functional governance committees are essential to be built and function even if various challenges need to be dealt with effectively during active conflict.

At the Strategic Layer, Ukraine’s digital sovereignty was and is continuously challenged by coordinated cyber activities such as espionage, disruption of communications, and disinformation campaigns that were at undermining national cohesion and strategic resilience in the complex hybrid warfare context of this conflict \cite{UseCase2}. Accordingly, by implementing clear sovereignty doctrines, aligned with national and international law, facilitates the transparent use of AI-driven analytics to rapidly identify, interpret, and mitigate disinformation threats \cite{UseCase3}. On this behalf, policies guided by ethical mandates and robust transparency ensures AI tools delivered interpretable outcomes, providing accountability which is crucial for maintaining legitimacy in both domestic and international arenas.

In the Governance Layer, structured oversight and verification processes such as system validation and red-teaming proved essential in Ukraine. Cyber security audits and immutable audit trails anchored via blockchain technologies ensure that cyber operations remained trustworthy and verifiable. Such governance mechanisms systematically evaluate and manage the risks posed by adversarial attacks, maintaining trust and enabling precise attribution of cyber incidents during operations \cite{UseCase1}. For instance, policy-driven autonomy requires clear, digitally embedded ROEs defining when AI can recommend or potentially initiate actions (e.g., blocking IPs, isolating networks) and when explicit human authorization is required.

The Data Sovereignty Layer is as important in safeguarding sensitive Ukrainian military data against unauthorized access and exfiltration during wartime cyber operations \cite{UseCase4}. This involves strict jurisdictional control where possible (e.g., processing sensitive data only on sovereign or vetted cloud instances within trusted regions) and employing technical measures like federated learning gateways for analysing shared threat intel without exporting raw logs, using data diodes for ingesting OSINT while preventing exfiltration, applying crypto-sharding or advanced encryption for data shared under specific agreements, and ensuring data minimization principles are followed despite the influx of information.

Within the AI Sovereignty Layer, it is  necessary to leverage secure MLOps pipelines and rigorous model validation methods, ensuring robustness against adversarial AI attacks. Federated learning and split-learning methodologies allow secure model updates without compromising data sovereignty, maintaining operational integrity despite relentless cyber threats. Moreover, the deployment of policy-aware models underpinned by transparent and explainable AI techniques reinforced ethical compliance and strategic alignment, significantly enhancing resilience during cyber operations \cite{UseCase5, UseCase6}. 

In the Cyber Operations and Infrastructure Layer, sovereign control is ensured over the actual execution of cyber defence actions based on AI insights. This involves managing defensive AI tools (anomaly detection, automated response recommendations), potentially supporting offensive cyber planning, and using simulations informed by agent-based modelling. To this end, Ukraine must maintain control over its trusted hardware and software stack, rigorously applying SCRM (Supply Chain Risk Management) processes even for donated equipment. Moreover, using national (military) clouds or secure enclaves (even if potentially hosted on allied/commercial infrastructure under specific agreements) forms the sovereign compute environment. Ensuring resilient network infrastructure control over military communications carrying AI data/commands is vital, alongside implementing zero-trust access principles to the digital platforms and ensuring Ukrainian commanders and other decision-makers retain ultimate authority via robust human override ("human on the loop") mechanisms. As control measures, for instance, operational autonomy necessitates resilient infrastructure that can function at least partially if allied connections are severed, and adaptive autonomy mechanisms are needed to adjust AI confidence thresholds based on the rapidly changing threat landscape and potential adversarial manipulation.

In this context, the risks encountered are multifaceted: direct destructive attacks (ransomware, wipers like CaddyWiper/Industroyer2), espionage, supply-chain attacks targeting donated systems, insider threats, and AI-specific vulnerabilities (e.g., data poisoning of shared intel feeds, model evasion) \cite{UseCase7, UseCase8}. Hence, risk mitigation requires a robust, layered defence (e.g., Zero Trust mechanism), proactive threat modelling and simulation specific to state-sponsored adversaries targeting Ukraine, resilient recovery plans various agencies, continuous monitoring and testing (e.g., adversarial AI testing), stringent supply chain vetting even for donation equipment, and strong cryptographic protection for both the data and models used \cite{UseCase10}. In addition to this, active threat monitoring and rapid incident response under sovereign control are critical operational requirements to be considered in this context. 

\section{Conclusions\label{section:discussion-conclusions}}
In today's evolving threat landscape, ensuring digital sovereignty has become mandatory for military organizations, especially given their increased development and investment in AI-driven cybersecurity solutions. This research has proposed a multi-angled framework to define and assess digital sovereign control of data and AI-based models for military cybersecurity. The framework focuses on context, autonomy, stakeholder involvement, and risk mitigation, aiming to protect sensitive defence assets.

The framework's layered architecture, ranging from strategic direction to operational execution, provides a comprehensive structure for establishing, assessing, and maintaining verifiable sovereign control over critical digital assets and AI-driven processes. The use case of the Ukraine conflict underscores the framework's applicability and the importance of maintaining operational autonomy, security, and ethical alignment in the face of sophisticated cyber threats.

Empirical validation of the proposed framework through pilot studies and real-world case applications is essential to refine the metrics for comprehensive assessment of digital sovereignty. Additionally, developing standardized protocols and collaborative frameworks would be beneficial for achieving a balanced approach to digital sovereignty, ensuring interoperability while respecting national and international interests.

\bibliographystyle{IEEEtran}
\bibliography{references}
\end{document}